\documentstyle[preprint,aps]{revtex}
\begin{document}
\draft
\title{Detecting the Photon-Photon Interaction by Colliding Laser Beam
Interferometry\\}
\author{M. Hossein Partovi\cite{email}}
\address{Department of Physics and Astronomy, California State University\\
Sacramento, California 95819-6041}
\date{\today}
\maketitle
\begin{abstract}
The feasibility of detecting the photon-photon interaction using Fabry-Perot
type laser interferometers developed for gravity wave
detection is demonstrated.  An ``external'' laser beam, serving as a
refractive
medium, is alternatively fed into
the cavities of the interferometer and made to collide with the ``internal''
beams thereby inducing a measurable phase difference between them.
[hep-ph/9308293]
\end{abstract}
\narrowtext
In 1933, Delbr\"uck \cite{one} suggested that quantum effects would cause
photons to be scattered by
an external electric field thereby violating the linearity of Maxwell
electrodynamics.  Shortly thereafter, Halpern \cite{2}, unaware of
Delbr\"uck's suggestion, realized that the process of pair creation would in
effect polarize the vacuum and give rise to a small photon-photon
interaction.
This interaction, he suggested, could in principle be observed in the
scattering
of light by light.
Delbr\"uck's phenomenon was first observed by R. Wilson \cite{3} in 1953 in
the scattering of 1.33 Mev gamma rays by the Coulomb field of the lead
nucleus.  Delbr\"uck scattering is essentially the interaction of a real
photon with a virtual one.  The interaction of two virtual photons, on the
other hand, is a common occurrence in present day high-energy collisions,
and has been studied extensively in recent years \cite{4}.

By contrast, the interaction of two real photons has never been
detected because of its extreme weakness.  The center-of-mass photon-photon
cross section is equal to $.031\alpha^{2}r_{e}^{2}(\omega/m)^{6}$ for
$\omega/m\ll1$ and nearly equal to it almost up to the pair production
threshold $\omega=m$ \cite{5}; here $\alpha$ is the fine-structure constant,
$r_{e}$ the classical electron radius, $\omega$ the photon energy, and $m$
the electron mass (natural units will be used until otherwise stated).
For 633 nm light, this cross section is about $4\times
10^{-64}$ cm$^2$, and it rises steeply to about $3\times
10^{-30}$ cm$^2$ near the production threshold.  These minute cross sections
are the reason why light-light scattering has not been much discussed in
connection with nonlinear electrodynamic effects arising from vacuum
polarization,
attention having been focused
instead on the magnetic counterpart of Delbr\"uck scattering, namely the
scattering of photons by an external magnetic field \cite{6}.  Nevertheless,
as the following discussion will show,
the combination of high intensity and frequency stability of
laser light, coupled with the astonishing sensitivity of present-day
Michelson
interferometers, makes it possible to overcome these incredibly small cross
sections and observe photon-photon interaction with real photons.

The purpose of this Letter is to demonstrate the feasibility of detecting
photon-photon interaction using the technology of Fabry-Perot type laser
interferometers developed for gravity wave detection \cite{7}.  The proposed
method relies on the fact that a beam of light essentially acts as a weakly
refractive medium for the propagation of another electromagnetic wave.
Therefore, if the output of an ``external'' laser is alternatively fed into
one of the two Fabry-Perot cavities of the interferometer at frequency $f$
and
made to collide with the ``internal'' beams, it will induce an
alternating relative phase difference of the same frequency between the
latter,
just as gravitational waves do in the case of gravity wave detectors.
To get an idea of the magnitudes involved, it is convenient to carry this
analogy
further.  For gravity waves, the quantity measured by the detector is the
{\it
gravitational-wave strain}, $h$, which is a measure of the deviation of
space-time curvature from its background value caused by the gravitational
wave.
For photon
interaction detection, the analogous quantity is the {\it refractivity},
$r$, of the
external laser beam inside the resonator cavities, defined as usual to be
the
deviation of the refractive index from unity.  The refractivity of a beam of
electromagnetic waves,
as shown by calculations presented below, is of the order
of $\alpha^{2}u/u_{e}$, where $u$ is
the energy density in the beam and $u_{e}=m^{4} = 1.4\times
10^{24}$
J/m$^{3}$ is essentially
the {\it Compton energy density of the
electron}.  In comparing the two cases, it must be remembered that there are
important differences between them favoring photon
interaction detection.  Among these, features that directly bear on the
question of
detection feasibility are (a) the fixed nature of cavity mirrors,
essentially limiting noise to photon shot noise, (b) the fact that the
signal frequency $f$ is
under total control, allowing longer photon detection times and other
fine-tuning
measures for sensitivity optimization, and (c) the enhancement of the
external
laser output (hence its refractivity) by a factor of $\sim
B^{2}$ in the resonator cavities, where $B$ is the mean number of
reflections in
the cavities.

In the following, we will use these considerations together with detailed
estimates to
show that sensitivity levels comparable to
those already achieved by prototype gravity wave detectors are sufficient to
detect photon-photon interaction by colliding laser beam interferometry
(CLBI).
Needless to say,
the detection and measurement of the interaction of light with light would
constitute an
important milestone not only as a direct test of fundamental quantum
electrodynamics but also as an ultrahigh-precision experiment.

The first step in the calculation is to find the interaction energy of an
assembly of photons to leading order in $\alpha$ and $\omega/m$.  As is well
known, the leading contribution to photon-photon scattering is of second
order in
$\alpha$ and originates in the box diagram in which four external photon
lines
are attached to a closed electron loop [8].  In leading order, the matrix
element for
this diagram is directly related to the interaction energy in question (in
much
the same way as the Born amplitude for potential scattering is related to
the
matrix element of the interaction potential).  Moreover, the $\omega/m$
($\sim
10^{-5}$ for visible light) limit of this contribution can conveniently be
cast in the
form of an effective Lagrangian modifying the Maxwell Lagrangian for free
photons, as was first shown by Euler and Heisenberg [9].

Using either the matrix element for the box diagram in the limit
$\omega/m\rightarrow 0$ or the Euler-Heisenberg Lagrangian, one finds, after
a
lengthy but
essentially straightforward calculation, the result [10]
\begin{eqnarray}
{\cal H}_{\text {int}} =&&-{\alpha^{2}\over 45m^{4}}{\sum_{{\bf e},{\bf
e'}}}\int \int [{\bf dk}][{\bf dk'}]n({\bf k},{\bf e}) n({\bf k'},{\bf e'})
\nonumber\\
&& \times
kk'{\Pi} ({\hat{\bf k}},{\bf e};{\hat{\bf k'}},{\bf e'})\eqnum{1}
\end{eqnarray}
for the desired interaction energy density.  Here $n({\bf k},{\bf e})$ is
the
number of photons of momentum ${\bf k}$ and (unit) polarization vector ${\bf
e}$,
$k=\vert {\bf k} \vert$, and $[{\bf dk]}$ stands for $d^{3}k/(2 \pi)^{3}$.
Moreover, ${\Pi}$ is a
non-negative, dimensionless number of order unity equal to $4{\cal R}^{2}+7
{\cal
I}^{2}$, where ${\cal R}$ and ${\cal I}$ are the real and imaginary parts of
the
quantity ${\bf f}{\cdot}{\bf f'}$, and ${\bf f} = {\bf e} + i {\hat{\bf
k}}\times
{\bf e}$.  Note that ${\cal H}_{\text {int}}$ represents a sum of negative
two-body interaction terms; {\it photons attract photons}.  Furthermore,
${\Pi}$
vanishes for ${\hat{\bf k}} = {\hat{\bf k'}}$ as a result of the vanishing
of the
center-of-mass energy of the pair; {\it parallel photons don't interact}.

For situations involving large numbers of photons, Eq. (1) can be
reformulated to
give the average first-order energy shift of a photon resulting from its
interaction
with the rest of the assembly.  Defining $\omega_{\text {int}}= \delta {\cal
H}_{\text {int}}/\delta n$, we find
\begin{equation}
\omega_{\text {int}}({\bf k},{\bf e})/\omega_{0} = -{2 \alpha^{2}\over
45m^{4}}{\sum_{\bf
e'}}\int [{\bf dk'}]{\cal E}({\bf k'},{\bf e'}){\Pi}
({\hat{\bf
k}},{\bf e};{\hat{\bf k'}},{\bf e'}),\eqnum{2}
\end{equation}
where ${\cal E}({\bf k},{\bf e}) = k n({\bf k},{\bf e})$ is the phase-space
energy
distribution function for the assembly, and $\omega_{0} = k$ is the energy
of a free
photon.  Note that the fractional energy shift $\omega_{\text
{int}}/\omega_{0}$
of a photon is independent of its (free) energy $k$ and only depends on the
orientations of its momentum and polarization.  This implies that the speed
of
propagation of electromagnetic waves through an assembly of photons is
reduced by
a factor of $1 - \omega_{\text {int}}/\omega_{0}$ relative to free space,
and
that therefore the photon assembly act as an anisotropic, linear,
refractive medium of refractivity $r = \omega_{\text {int}}({\bf k},{\bf
e})/\omega_{0}$ [11].  Eq. (2) can thus be used to find the refractive index
for
any desired configuration of interacting electromagnetic waves.

The configurations of interest here are those involving {\it beams}, for
which
${\cal E}({\bf k'},{\bf e'})$ would be concentrated along some fixed
direction, say ${\hat{\bf k}_{0}^{\prime}}$.  The corresponding refractivity
is then found to be
\begin{equation}
r({\hat{\bf k}},{\bf e};{\hat{\bf k}_{0}^{\prime}}) = {2 \alpha^{2}\over 45}
(u/u_{e}) (1-\cos \theta )^{2} [4 + 3 \langle
\sin^{2}(\psi - \phi )\rangle ].\eqnum{3}
\end{equation}
Here $u = {\sum_{\bf
e'}}\int [{\bf dk'}] {\cal E}({\bf k'},{\bf e'})$
is the energy density of
photons in the refracting medium,
$(\theta , \phi , \psi )$ are the three Euler angles that
rotate the $({\bf e}, {\hat{\bf k}} \times {\bf e}, {\hat{\bf k}})$ triad
onto
$({\bf e'}, {\hat{\bf k}_{0}^{\prime}}\times {\bf e'}, {\hat{\bf
k}_{0}^{\prime}}
)$, and angular
brackets are used to denote the polarization average with respect to ${\cal
E}({\bf k'},{\bf
e'})$.  Recall that ${\hat{\bf k}_{0}^{\prime}}$ is now the (fixed)
direction of the refracting beam.
Note also that $\cos \theta = 1(-1)$ corresponds to parallel (antiparallel)
beams, confirming the absence of interaction for parallel photons.  For
colliding
(i.e., antiparallel) beams, Eq. (3) reduces to
\begin{equation}
r({\bf e}) = {8 \alpha^{2}\over 45}
(u/u_{e}) [4 + 3 \langle {\vert {\bf e} \times {\bf e' \vert}^{2}}\rangle ].
\eqnum{4}
\end{equation}
Note that the polarization average in (4) is zero for parallel-polarized
beams, reaches
a maximum of unity for cross-polarized beams,
and reduces to a half if either beam is unpolarized.

It is also worth noting here that if the photons of the refracting medium
are
unpolarized (but
otherwise of arbitrary configuration), Eq. (2) reduces to
\begin{equation}
r({\hat{\bf k}}) = {11 \alpha^{2}\over 45 m^{4}} \int [{\bf dk'}]
(1-{\hat{\bf k}}
\cdot {\hat{\bf k'}})^{2}{\cal E}({\bf k'}), \eqnum{5}
\end{equation}
where ${\cal E}({\bf k'})$ is the polarization-summed energy distribution
function of the refracting medium.  One can verify that Eq. (5) agrees with
Eq. (4) for the case of antiparallel beams at least one of which is
unpolarized.

We are now in a position to derive detection limit estimates for CLBI.  As
mentioned at the outset, the configuration considered here is that of the
basic Fabry-Perot type interferometer used for gravity wave detection [7]
suitably
modified to accept the output of an external laser into its cavities [12].
Inside each cavity, the external laser beam acts as a refractive medium,
colliding with the internal beam and inducing in it a phase change
proportional to the refractivity of the external beam.  The two
lasers must be operated at (or very near, depending on the detailed mode
in which the interferometer is operated) the resonance frequencies of the
cavities, and the two frequencies must be sufficiently different to allow an
efficient filtering
of the external beam photons so that only internal beam photons find
their way to
the photodetector. Furthermore, the input of the external laser into the
cavities can be modulated so as to alternate between the two with a
frequency
$f$, say according to the modulation factors $[1 \pm \cos ( 2 \pi f t )]/2$.

Let the power output of the internal and external lasers be
$P_{\text {int}}$ and $P_{\text {ext}}$, respectively.  The output of the
lasers will be assumed to be unpolarized, so that the refractivity of the
external beam as given by Eq. (4) or (5) is equal to ${44\over
45}(u/u_{e})$.
Therefore the amplitude
of the difference in the refractivities of the external beams inside the two
cavities, denoted by $ \Delta r_{\text {ext}}$, is, up to factors of order
unity, given by
$\alpha^{2} B^{2} P_{\text {ext}}/c u_{e} A$, where $A$ is the average
cross section of the external beam
inside the cavities and $c$ is the speed of light in vacuum (SI units
will be used hereafter).  Recall that $B$ is the mean number of reflections
inside the cavities ($\sim$ cavity finesse).  Thus with $L$ representing
the length of each cavity, the amplitude of the relative phase between
the two internal beams arising from the difference in the refractivities
is given by the usual formula $BL \Delta r_{\text {ext}}/ \lambdabar $,
where
$\lambdabar$ is the reduced wavelength of the internal laser
light.

A customary and useful way of characterizing the detection limit of the
interferometer, assumed to be governed by the photon shot noise, would be
to define a {\it shot-noise equivalent} refractivity
\begin{equation}
r_{\text {shot}} \approx \left[ {\hbar c \lambdabar f \over P_{\text {int}}
(BL)^{2}}\right] ^{1/2}, \eqnum{6}
\end{equation}
which (up to factors of order unity) is the amplitude of the refractivity
difference that would produce the same signal at the photodetector as the
shot noise
[7,13].  It should be noted here that Eq. (6) assumes that (a) the
photodetector integration time $\tau$ is matched to the modulation
frequency $f$ for optimal sensitivity, $\tau = (2f)^{-1}$, and (b) $\tau$
is not less than the mean life time of an internal beam photon in the
cavities, $ \tau \geq BL/c$.

Using Eq. (6) and the expression for $r_{\text {ext}}$, we arrive at
an estimate of the signal-to-noise ratio for CLBI:
\begin{equation}
\Delta r_{\text {ext}} \div r_{\text {shot}} \approx {\alpha^{2} B^{2}
P_{\text {ext}} \over c u_{e} A} \div \left[ {\hbar c \lambdabar f
\over P_{\text {int}} (BL)^{2} }\right] ^{1/2}.
\eqnum{7}
\end{equation}
A more convenient form of Eq. (7) for numerical estimates is
\begin{eqnarray}
\Delta r_{\text {ext}}/r_{\text {shot}} \approx &&25
{L \over 10{\text{m}}}{10^{-5}{\text{m}}^{2}\over A}{P_{\text {ext}}
\over 50{\text{W}}} \left( {B \over 10^{4}} \right)^{3}\nonumber\\
&& \times
\left( {P_{\text {int}}
\over 50 {\text{W}}}{10^{-3}{\text{Hz}} \over f}{10^{-7}{\text{m}} \over
\lambdabar } \right)^{1/2}. \eqnum{8}
\end{eqnarray}
This formula can be used to estimate the expected detection limit of a
colliding laser beam interferometer.  The individual magnitudes of
$\Delta r_{\text {ext}}$ and $r_{\text {shot}}$ for the numerical
values used in Eq. (8) are $6.3 \times 10^{-23}$ and
$2.5 \times 10^{-24}$, respectively.  Note that the fiducial
numerical values used in Eq. (8) are consistent with the
present-day technology of gravity
wave detectors.  Indeed some of these values (e.g., $f$ or $L$) can be
extended in the
direction of boosting the ratio in Eq. (8) without much difficulty.

The estimate given in Eq. (8) and the arguments leading to
it provide the evidence for the feasibility of detecting the interaction of
light with light by means of CLBI.  Compared to gravity wave detection,
CLBI enjoys three important advantages mentioned before, namely the
absence of moving parts, the total controllability of the signal frequency,
and the possibility of signal amplification by means of external laser
intensity build-up inside the cavities (cf. the $B^{2}$ dependence in
$\Delta r_{\text {ext}}$).  On the other hand, CLBI requires the
design and development of the extra optics required for piping,
aligning, focusing, and filtering the colliding laser beams.  Incidentally,
a numerically accurate version of the estimates given above can only be
given when details of the optics such as the transverse profiles
of the colliding beams inside the resonators are specified.

We conclude by recalling that our aim in the above discussion
has been to establish the essential feasibility of detecting the
photon-photon interaction with colliding laser beams by means of a
detailed analysis of a specific experimental arrangement.  This arrangement
is really a minimal and generic modification of Fabry-Perot type
interferometers developed for gravity wave detection
and may not be the most suitable design for carrying out the actual
experiment [14].  Nevertheless, it does serve to establish that the phenomenon
in question is within present experimental reach.

This work was supported in part by a research award from the California
State University, Sacramento.


\begin{references}
\bibitem{11} That photon-photon interactions give rise to anisotropic
refractivity and other exotic effects has been known for a long time; see
Ref.
[6].
\bibitem{13} In general $B$ depends on wavelength and need not be the same
for
the two light sources used in the interferometer.  However, we are only
interested in estimates here and will therefore overlook the difference.
\bibitem{14} For example, we have not exploited the polarization dependence
of the interaction at all.  As made clear by remarks following Eq. (4), the
refractivity of a beam has a strong polarization dependence which can be
used to advantage in the optical processing of the beams.
\end{references}
\end{document}